# Assessing Effectiveness of Pulsed Input on Mixing Characteristics of Non-Newtonian fluids in T-shaped Channels.

Anirban Roy[1], Avinash Kumar[1], Chirodeep Bakli[1] and Gargi Das[2]

[1]Thermofluidics and Nanotechnology for Sustainable Energy Systems Laboratory, School of Energy Science and Engineering, Indian Institute of Technology Kharagpur, Kharagpur-721302, India

[2]Multiphase flow Laboratory, Department of Chemical Engineering, Indian Institute of Technology Kharagpur, Kharagpur-721302, India

**ABSTRACT**

Mixing of reagents in microfluidics is necessary for various applications however due to the laminar nature of flows, efficient mixing in a small span of length and time becomes difficult. The analysis of mixing of non-Newtonian fluids is critical as they are commonly encountered in practical applications. Towards this, we investigated an effective way for mixing of non-Newtonian fluids using pulsatile velocity inlet conditions. In the present study, the non-Newtonian fluid is modelled using the power law model with varying fluid rheology from shear thickening to shear thinning. For enhancing the mixing, pulsed velocity inlet condition is applied with varying phase angle and frequency and compared with constant velocity inlet condition. We demonstrated enhanced mixing using pulsing velocity inlet condition and achieved a maximum mixing of 97.6% using pulsed input velocity with a phase difference of 180° and considering a frequency of 5 Hz for the case of shear thinning fluid (n=0.6). For the same condition, the mixing index is 89.1% and 85.2% for Newtonian and shear thickening fluid (n=1.4), respectively. The present study will be helpful in designing micromixers for mixing non-Newtonian fluid effectively in a small span of length and time.

## 1. INTRODUCTION

Microfluidics is the science of systems that process or manipulate fluid flow through microchannels. Owing to the advantages [1] like precise controllability, excellent operating stability, and enhanced transport properties, microfluidic devices have been used in various applications ranging from small fuel cells and molecular diagnostics [2], to 3D printing [3], electronics cooling [4,5] and micro-mixing.

The mixing of two reagents is often a crucial stage in a procedure, regardless of the application of microfluidics. However, it remains a challenge in microchannels owing to laminar nature of flow. Many researchers have utilized techniques like geometry variation [6,7], electric and ultrasonic vibratory fields [8], and time dependent inlet conditions [2,9,10] to enhance the mixing process in micro domain. Out of these, the time dependent inlet condition has gained massive attention owing to its simplicity in practical applications. Fluids can be broadly classified into two categories- Newtonian fluids and non-Newtonian fluids. Since most of fluids encountered in day-to-day applications are non-Newtonian in nature, it becomes important to categorically study the mixing of such fluids in micro domain as well. In this study, an attempt is made to evaluate the mixing characteristics of non-Newtonian fluids, particularly fluids that follow the power law under the influence of time-varying velocity input.

## 2. LITERATURE REVIEW AND OBJECTIVE

Most intuitive way of increasing mixing in microchannels is by introducing obstructions in the channel which generate vortices to enhance mixing. This concept has been studied extensively by many researchers [11–17] and have proved to increase the mixing by significant amount. However, fabrication of these special geometries in micro domain becomes complicated due to their smaller size and required accuracy and precision. As a result of this, a relatively simpler method-pulsed velocity inlet- caught attention of many researchers.

The first instance of mixing enhancement by the application of time varying velocity at inlet was done by Glasgow and Aubry [2,9]. They studied the effect of various combination of pulsing and non-pulsing sinusoidal inlet velocity condition on mixing of two different concentration of water. They found the mixing time to reduce drastically when compared to steady flow inlet. They also found the optimal mixing condition occurs when the phase angle between the two pulsed input is 180°. The effect of microchannel geometry on mixing in pulsed inlet conditions was studied by Goullet et al. [8]. They found that effect of only pulsing the input results in better mixing than that obtained by geometric variations in the microchannel. The combination of both results in substantial increase in mixing of two fluids. An increase of mixing index by 19% was reported by Afzal and Kim [10] where they used pulsed input along with convergent-divergent geometry which is sinusoidal in nature.

From the literature reviewed, it has been observed that significant amount of work has been done in analyzing the mixing characteristics of Newtonian fluids by the use of both geometric variations and pulsed velocity inlet. The mixing analysis of non-Newtonian fluids, however, has been limited to geometric variations only.

In the present study, we evaluate the mixing characteristics of non-Newtonian fluids in simple T shaped geometry with perpendicular inlets and with pulsed velocity input. The effect of frequency and the phase angle on mixing is also evaluated. Finally, an optimal frequency and inlet condition for shear thickening and shear thinning fluid is also suggested.

## 3. METHODOLOGY

### 3.1 Geometry

A T shaped micro-channel with two inlets perpendicular to each other was used in our analysis. Domain of analysis with various dimensions is shown in the Figure 1.

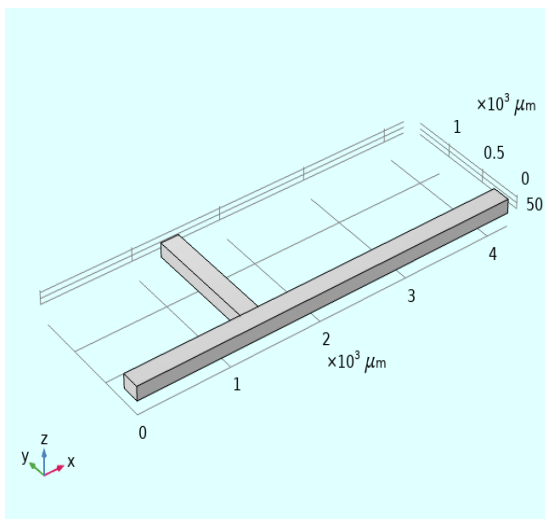
(a)

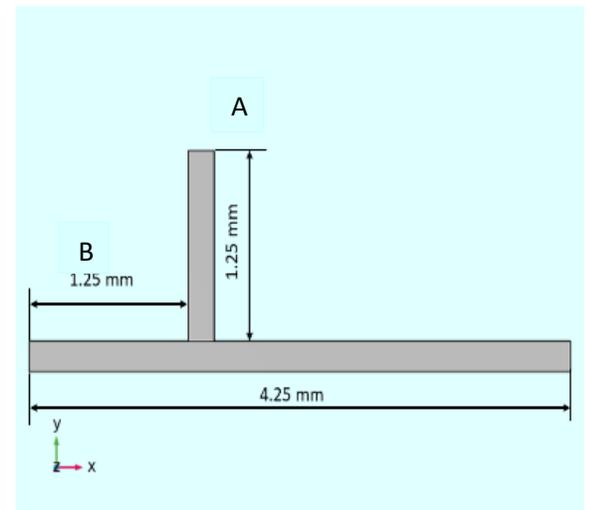
(b)

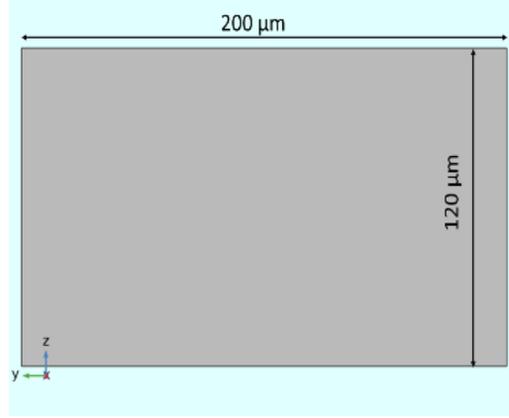

**(c)**

**Figure 1:** Analysis domain (a) Isometric view (b) Top view (c) Cross-section

The inlet and outlet arms are of 1.25 mm length while the main channel after confluence is of 3 mm length. All the channels have same cross-sectional area of (200 X 120) µm.

### 3.2 Numerical Modeling

A 3D numerical study was performed and the governing equations for modelling of the physics involved are as follows-

Continuity equation

$$\nabla \cdot \mathbf{u} = 0 \tag{1}$$

Momentum equation

$$\frac{\partial \mathbf{u}}{\partial t} + (\mathbf{u} \cdot \nabla)\mathbf{u} = \frac{1}{\rho}\left[-\nabla p + \nabla \cdot \{2\mu_{app}(\Gamma)\Gamma\}\right] \tag{2}$$

Where,

$$\Gamma = \frac{\left(\nabla \mathbf{u} + (\nabla \mathbf{u})^T\right)}{2} \tag{3}$$

$$\mu_{app}(\Gamma) = m(2\Gamma)^{n-1} \tag{4}$$

Species transport equation

$$\frac{\partial c}{\partial t}+(\mathbf{u}.\nabla)c = D\,\nabla^2 c \tag{5}$$

These equations were solved using COMSOL Multiphysics which is a FEM based solver.

At inlet, the applied average velocity was assumed to sinusoidally vary as a combination of steady state velocity and the fluctuating component. The equation of velocity at inlet A is $1.0+7.5\sin(2\pi ft)$ mm/s while that of inlet B is $1.0+7.5\sin(2\pi ft+\varphi)$ mm/s.

### 3.3 Validation study

The numerical model was validated with the results published by Glasgow and Aubry [2] and the results were found to be within the acceptable range of error. The comparison of the contour plots is shown in Figure 2.

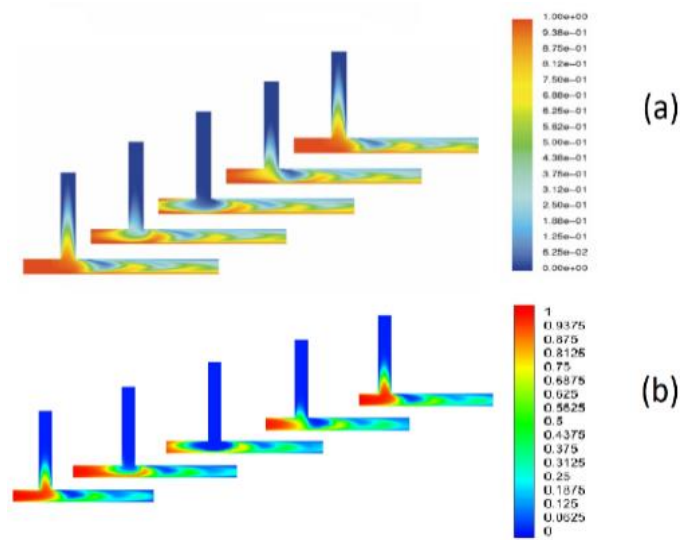

**Figure 2:** (a) Glasgow and Aubry [2] (b) Current Simulation

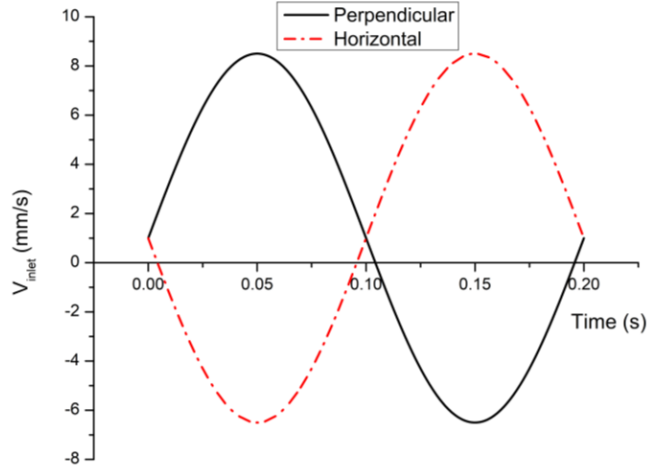

**Figure 3:** Inlet velocity profiles used for validation

The readings are taken after the pulsing is fully established and at an interval as shown in Figure 3.

Study of non-Newtonian fluids with different flow behaviour index (n) 0.6,1 and 1.4 have been done. n<1 represents the behaviour of shear thinning (pseudoplastic) fluid, and n<1 represents shear thickening (dilatant) fluid. n=1 represents Newtonian fluids and this is used as a base line for comparison of certain behaviour. The flow consistency index (m) is kept fixed at 0.001 Pa.s while the density of the fluid is taken as 1000 kg/m$^3$.

### 3.4 Evaluation of mixing index

In order to quantify mixing of fluid, a parameter called mixing index or mixing efficiency is defined. Mathematical formulation of the mixing index is shown in Equation 6.[14]

$$Mixing\ Index = \left(1 - \sqrt{\frac{\sigma^2}{\sigma_{max}^2}}\right) \qquad (6)$$

Where,

$$\sigma^2 = \frac{1}{N}\sum_{i=1}^{N}\left(c_i - c_m\right)^2 \tag{7}$$

Here, the $\sigma^2$ is the actual variance and $\sigma^2_{max}$ denotes the maximum possible variance of concentration of the fluid. $c_i$ is the concentration at the sampling point 'i'. N is the total no. of sampling points over a surface.

## 4. RESULTS AND DISCUSSION

In this section, we present our observations and some key points in our analysis. We start with mesh independence study followed by the effect of pulsed input, phase angle and pulsing frequency, on the mixing characteristics of non-Newtonian fluids with the properties mentioned earlier. The inlet mass fraction at perpendicular inlet is taken as 0 while at horizontal inlet is taken as 1. All the readings are taken when the pulsing is fully established i.e., at least after 10 periods. The readings are taken at a location 0.75 mm after confluence.

### 4.1 Mesh independence study

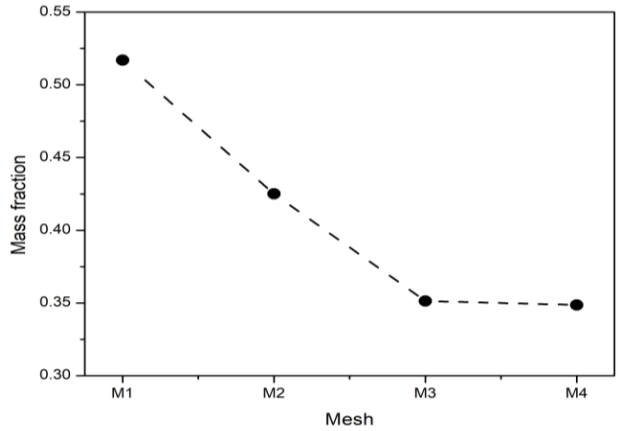

**Figure 4** Mesh independence study

Mesh independence study was performed with the meshes with no. of elements mentioned in Table 1. It was observed that after mesh M3, the variation in mass fraction of the fluid was negligible. It can be also be observed in the Figure 4. The computational time however increased significantly for the M4 mesh. Therefore, we can safely assume that the M3 is the optimal mesh for our study.

**Table 1:** Mesh nomenclature

| Mesh elements | Name |
|---|---|
| 64043 | M1 |
| 117375 | M2 |
| 304542 | M3 |
| 1024680 | M4 |

## 4.2 Effect of pulsed input

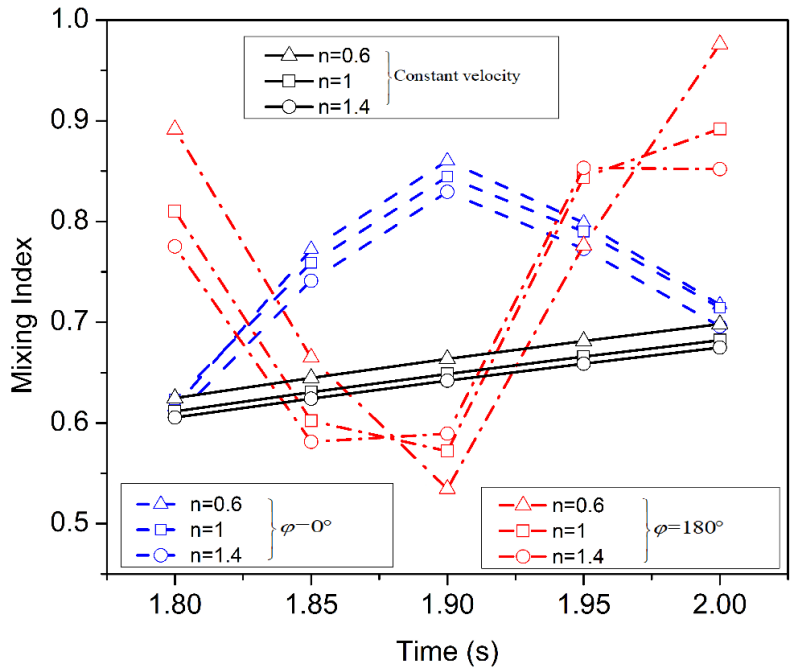

**Figure 5:** Effect of phase angle on mixing for all the fluids.

We begin with comparing the mixing of various non-Newtonian fluids with varying flow behaviour index, n from 0.6 – 1.4, in a T shape microchannel with pulsed and without pulsed input. For the case of pulsed flow, the results have been taken when pulsed flow in fully established (for present study from 1.8 s - 2 s). It is observed from figure 5 that the mixing index is higher for shear thinning fluid and lower for shear thickening fluid than the Newtonian fluid regardless of input condition (pulsed or constant velocity input). This is because for the former case, with decrease in flow behaviour index, the apparent viscosity decreases which increases the flow velocity and consequently mixing enhances. Whereas for the latter case, with increasing flow behaviour index, the apparent viscosity increases and thereby reduces the mixing. It can be noted that, for the case of constant input velocity, the mixing enhances linearly with time. However, for the case of pulsed input condition, the mixing enhances, attains maximum mixing, at half cycle; and beyond that, it decreases with time. For the case of constant input flow, the mixing index is 69.8%, 68.2%. 67.5% for n = 0.6, n=1 and n=1.4, respectively whereas for the case of pulsed flow, the mixing index is 71.7 %, 71.4% and 69.5% for n = 0.6, n=1 and n=1.4, respectively after full cycle, (t=2 s). However, for the case of pulsed input, the maximum input index is obtained at half cycle with 86%, 84.4%, and 82.9% for n = 0.6, n=1 and n=1.4 respectively.

**4.3 Effect of phase angle**

In this section, the effect of phase angle on the mixing index is studied for all the fluid rheology. We have compared the input condition of 180° phase difference with standard pulsing case ($\varphi=0°$) considering the frequency to be 5 Hz. It is observed from figure 5 that, for the case of $\varphi=180°$, unlike the standard pulsing case, the mixing index reduces with time up to half cycle and increase beyond that regardless of fluid rheology. This is because for the first half cycle (1.8 s – 1.9 s), as time increases, the fluid is being pushed in from one inlet while being pulled out from other inlet; this results in filling in of the channel with only single fluid. Then, for second half cycle (1.9 s – 2 s) the fluid from first inlet is pulled out and from second inlet is being pulled in and as the second fluid is being pushed in, it is getting diffused with the fluid pushed in, in the first half cycle creating a recirculating flow at the junction and thereby enhancing the mixing. It can be noted that, for the case of $\varphi=180°$, the maximum mixing is achieved after full cycle with the mixing index, of 97.6%, 89.1% and 85.2% for n = 0.6, n=1 and n=1.4 respectively.

**4.4 Effect of frequency of pulsation**

As discussed in the aforementioned section, the highest mixing is achieved for pulsed input condition of 180° phase difference. Therefore, considering the phase difference of 180 °, we varied the frequency of pulsation ranging from 5Hz to 20Hz and investigated its impact on the mixing of different fluid rheology. It is observed from figure 6 that with increasing frequency the mixing index decreases regardless of fluid rheology. This is attributed to the fact that with increasing frequency, the full cycle time decreases due to increase in the number of cycles for a fixed range of time (1.8 s to 2 s) and thereby fluids do not get enough time to diffuse and consequently decreases mixing. Also, it is observed that, the pattern of variation of mixing index with time changes as the frequency increases. This is attributed to increase in number of cycles for a fixed range of time. It is interesting to note that the mixing index reduces significantly from 5 Hz to 10 Hz. However, the variations in the same in minor beyond 10 Hz.

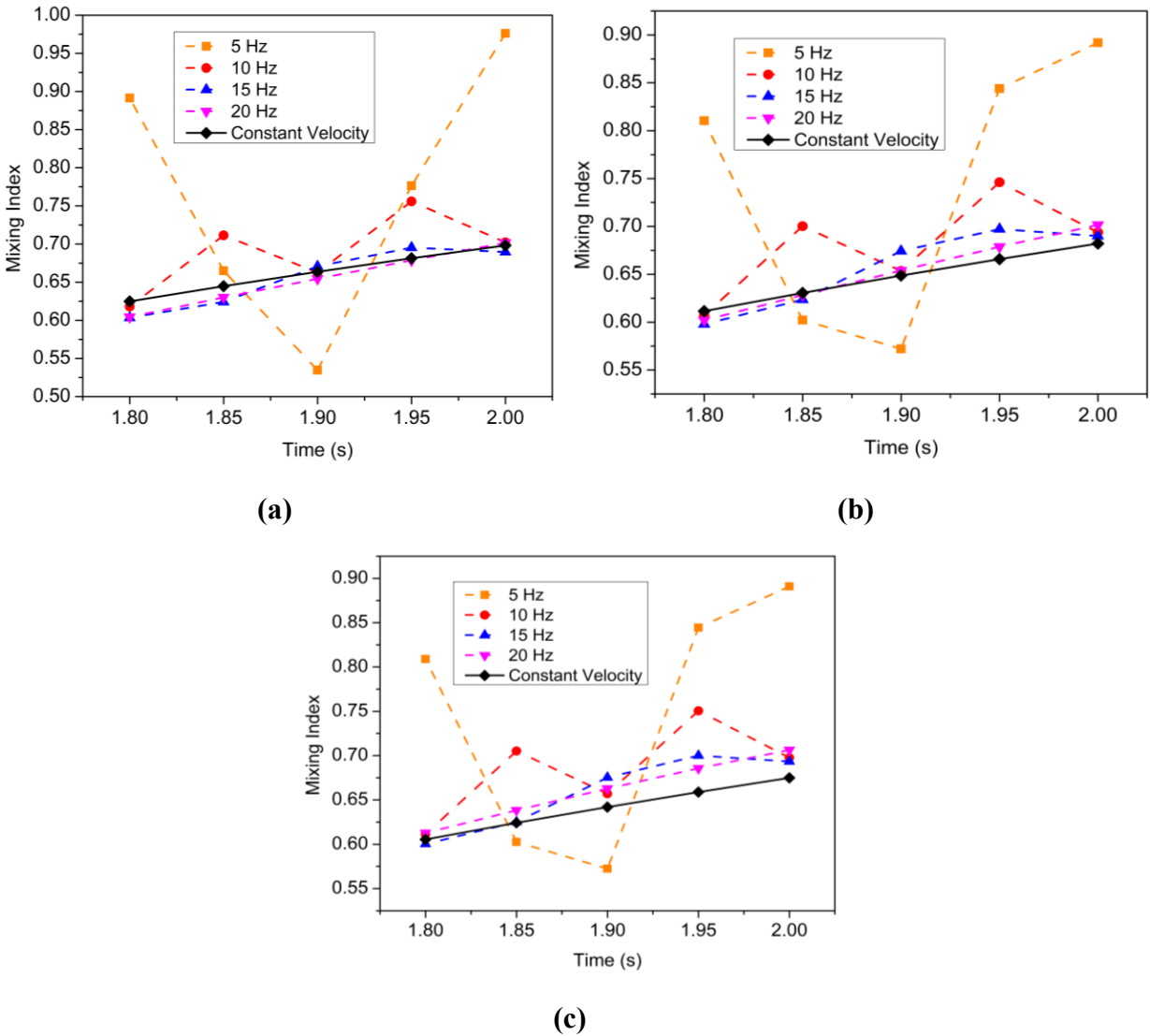

**Figure 6:** Effect of frequency on mixing index of (a) shear thinning fluid (n=0.6) (b) Newtonian fluid (n=1) (c) shear thickening fluid (n=1.4).

## 5. CONCLUSIONS

We have investigated numerically the fluid mixing in a T shaped channel with different flow inlet conditions, constant velocity, standard pulsing case (φ=0°) and pulsing inlet with 180° phase difference for various non-Newtonian fluids. We demonstrated that, the mixing of fluid increases with the decreasing flow behaviour index. Also, it is evident that, the standard pulsed flow improves the mixing marginally than the constant velocity for full cycle. However, for the case of half cycle, the standard pulsed flow gives maximum mixing index, 86%, 84.4%, and 82.9% for n = 0.6, n=1 and n=1.4 respectively. Next, we achieve the highest mixing by increasing the phase difference of pulsed input by 180° considering frequency of 5 Hz with the mixing index of 97.6%. Lastly, we investigated the effect of frequency on mixing index and found that with increase in frequency, the mixing index reduces. Finally, it can be concluded that, for enhancing mixing in power law fluids, the pulsatile inlet condition along with lower frequency is an effective technique and best mixing can be achieved with phase difference of 180°. The present study will be helpful in designing micromixers for better and effective fluid mixing in small time and span of channel.

**NOMENCLATURE**

| | | |
|---|---|---|
| $\boldsymbol{u}$ | Velocity vector | [m/s] |
| $t$ | Time | [s] |
| $p$ | Pressure | Pa |
| $\mu_{app}$ | Apparent viscosity | [Pa.s] |
| $\rho$ | Density of fluid | [kg/m$^3$] |
| $\boldsymbol{\Gamma}$ | Rate of strain tensor | [1/s] |
| $\Gamma$ | Magnitude of $\boldsymbol{\Gamma}$ | [1/s] |
| $m$ | Flow consistency index | [Pa.s] |
| $n$ | Flow behaviour index | [---] |
| $c$ | Concentration | [mol/m$^3$] |
| $D$ | Diffusion coefficient | [m$^2$/s] |
| $f$ | Frequency | [Hz] |
| $\varphi$ | Phase angle | [°] |